\documentclass[aps,twocolumn,nofootinbib,superscriptaddress]{revtex4-1}
\usepackage{amsmath}
\usepackage{psfrag}
\usepackage{graphicx}
\usepackage{dcolumn}
\usepackage{units}
\usepackage{tikz,pgfplots}
\usepackage{pgfplotstable}
\newcommand{\marker}{}
\newcommand{\DB}{}

\begin{document}
\title{Longitudinal rms emittance preservation during adiabatic capture for Gaussian beams}
\author{Denys Bast}
\affiliation{Technische Universit\"at Darmstadt, Institut f\"ur Teilchenbeschleunigung und Elektromagnetische Felder, Schlo\ss gartenstra\ss e 8, D-64289 Darmstadt, Germany}
\author{Uta Hartel}
\affiliation{GSI Helmholtzzentrum f\"ur Schwerionenforschung GmbH, Planckstra\ss e 1, D-64291 Darmstadt, Germany}
\author{Harald Klingbeil}
\affiliation{Technische Universit\"at Darmstadt, Institut f\"ur Teilchenbeschleunigung und Elektromagnetische Felder, Schlo\ss gartenstra\ss e 8, D-64289 Darmstadt, Germany}
\affiliation{GSI Helmholtzzentrum f\"ur Schwerionenforschung GmbH, Planckstra\ss e 1, D-64291 Darmstadt, Germany}
\date{\today}

\begin{abstract}
This paper presents an analysis of the emittance preservation during a heavy ion bunching process. This analysis aims at finding the correlation between the rms emittance of a coasting beam and that of the resulting bunches. The emittance of a Gaussian coasting beam bunched adiabatically in synchrotrons is investigated using theoretical considerations and particle tracking simulations. New results are presented on the relation between the emittance before and after the bunching process. It is found that a constant factor can be determined to describe this relation.

This factor can be used to avoid time-consuming simulations. If the rms emittance is known before the bunching process, it can be calculated for the bunched beam afterwards. This also makes it possible to determine the required bucket area for a certain filling ratio.
\end{abstract}

\maketitle
\section{Introduction}
\label{sec_intro}
Before the acceleration process of ions in synchrotrons such as SIS18 at GSI, coasting beams have to be bunched. This bunching process happens by applying an RF \DB{voltage} when no acceleration takes place yet. 
Liouville's theorem guarantees phase space area preservation for time-dependent Hamiltonian systems (cf.~\cite{lee, garoby, Klingbeil2015, ruggiero}). As soon as the RF amplitude is turned on or changed, Liouville's theorem is no longer strictly applicable since the orbits of the particles in phase space are not closed anymore. The change of the RF amplitude leads to emittance growth in general.
Furthermore, the area preservation is only guaranteed for a continuous particle density with clear boundaries in phase space. The theorem does not hold for a discrete distribution of individual particles \cite{ruggiero}.

The concept of adiabaticity indicates the possibility to limit the emittance growth by increasing the RF voltage slowly enough. 
A common approach is to define an adiabaticity parameter with respect to the synchrotron frequency and perform studies about how slowly the voltage has to be increased to limit the emittance growth to a certain limit (cf. \cite{Ng2012, Mohite11}). In principle, it is possible to reduce the growth to any given percentage by making the process slow enough \cite{Klingbeil2015}.

The starting point of this paper is to thoroughly analyze the emittance of coasting beams and that of bunched beams. If the bunching process is slow enough, i.e.~perfectly adiabatic, there should be a fixed ratio of one between the emittance before and after the adiabatic process.
First, we present theoretical considerations on the longitudinal emittance and present our approach for describing the relationship of the rms emittance before and after bunching. Starting with the parameters of SIS18, simulations of longitudinal particle dynamics are carried out. The parameters are then extended to other possible accelerator configurations.
Thereby we determine a correlation between the rms emittance of a coasting beam before the adiabatic bunching process and the rms emittance of the bunched beam. 

\subsection{Longitudinal emittance}
\label{subsec_longEmittance}
The longitudinal emittance of an ion beam is defined by the area in the longitudinal phase space that is occupied by the ions. The state of every ion at a time $t$ is characterized by its energy deviation $\Delta W$ and time deviation $\Delta t$ with respect to the so-called reference particle. Because by definition the reference particle is always in compliance with the desired energy, revolution time and all other target parameters, the reference particle corresponds to the center of the phase space $(\Delta W,\Delta t)=(0,0)$. \marker{If the particles are uniformly distributed along the whole accelerator circumference, the beam is called  a coasting beam. A bunched beam is obtained, if the particles are focused in longitudinal direction. In case of a bunched beam, the particles oscillate around the reference particle according to the phase focusing principle (cf.~\cite{lee,wiedemann,hinterberger}). In a single-harmonic RF bucket\footnote{\marker{without small angle approximation}} it is known that the closer the particles are to the reference particle, the higher is their oscillation frequency.}

We first assume a continuous uniform distribution of the particles. This first approach makes it easy to calculate the occupied area in phase space. All borders are clearly defined and thus for both, a coasting beam and a bunched beam, \marker{a longitudinal emittance can be determined}.

In the next step, a Gaussian distribution is assumed as a more realistic assumption for a coasting beam. Neglecting space-charge effects, also the bunched beam can approximately be described by a 2D Gaussian distribution, as confirmed by our simulations and by literature (cf.~\cite{reiser}). A Gaussian shape is obtained for both phase space dimensions and for the projection onto each axis (as long as the Gaussian distribution is narrow enough to be cut off close to the separatrix).

In case of a beam with discrete particles, defining an area leads to the necessity of further assumptions, since exact limits for the area are not given. In the following sections commonly used rms emittance definitions are presented and discussed in order to motivate the aim of the further investigations. The adiabatic bunching has been chosen as a \marker{fundamental process (cf.~\cite{garoby,cottingham}), which is needed for the standard operation in synchrotrons such as SIS18 at GSI e.g.~before acceleration or for de-/re-bunching the beam at flattop energy}. No higher-order effects like beam loading, space charge or impedances are taken into account.

\subsection{Coasting beam}\label{subsec_coasting_beam}
If no amplitude of the accelerating RF signal or gap voltage\footnote{\marker{The gap voltage denotes the accelerating RF voltage induced in a ceramic gap, which interrupts the metallic beam pipe of a cavity.}} is present, the ions travel equally distributed along the circumference of the synchrotron. Hence, the ions of the coasting beam have an average revolution time $T_R$ in the accelerator\footnote{\marker{The index $R$ marks quantities which belong to the reference particle.}}. The longitudinal emittance of a uniformly distributed continuous coasting beam is therefore obtained by
\begin{align}
A_{b,coast}^{unif} = 2\cdot \Delta \hat{W}\cdot T_R\,.\label{eq_A_b,coast}
\end{align}
Here $\Delta \hat{W}$ denotes the energy deviation of the outermost particle in the coasting beam. The factor 2 is due to the $\pm\Delta W$ band in phase space.

Assuming a Gaussian distribution of a coasting beam, often the $2\,\sigma$ value of the momentum spread $\frac{\Delta p}{p_R}\big|_{2\sigma}$ or the energy spread $\frac{\Delta W}{W_R}\big|_{2\sigma}$, within which about 95\% of the particles are located, is used. \marker{Here, $\Delta p$ and $\Delta W$ denote the momentum deviation and the energy deviation with respect to the reference particle with the momentum $p_R$ and the total energy $W_R$.} 
The underlying assumption that the emittance always corresponds to the area that encloses 95\% of the particles, however, does not necessarily lead to emittance preservation as we will show. Therefore, we choose a more general emittance approach taking Liouville's theorem into account introducing a not yet known factor $k$. \marker{This factor $k$ defines an effective outer limit of the energy deviation of a continuous beam that is regarded equivalent to the Gaussian beam with respect to the calculation of the emittance.} We define the sigma emittance of the continuous coasting beam with Gaussian momentum distribution by
\begin{align}
A_{b,coast}^{Gauss} &= 2\cdot k\cdot\sigma_{\Delta W,coast}\cdot T_R\,,\label{eq_A_b,coast_sigma}
\end{align}
\marker{if the $\sigma$ value of the energy spread $\sigma_{\Delta W,coast}$ is used. In this paper, the $\sigma$ quantities describe the continuous Gaussian distribution. In the case of a discrete distribution, e.g.~a sufficiently large data set generated by a simulation or by the finite number of particles, we define the rms emittance of a coasting beam by}
\begin{align}
A_{b,coast} &= 2\cdot k\cdot\Delta W_{rms,coast}\cdot T_R\,.\label{eq_A_b,coast_rms}
\end{align}
\marker{Here, the rms quantity $\Delta W_{rms}$ of $N$ particles \marker{with the energy deviation $\Delta W_i$ per particle $i$} according to
\begin{align}
\overline{\Delta W} &= \frac{1}{N}\sum\limits_{i=1}^N \Delta W_i \label{eq_Wmean}\\
\Delta W_{rms} &=\sqrt{\frac{1}{N}\sum\limits_{i=1}^N \left(\Delta W_i-\overline{\Delta W}\right)^2}\label{eq_Wrms}
\end{align}
is \marker{applied}. These rms quantities can be calculated independent of the distribution. Only if the discrete data belong to a Gaussian distribution and if $N$ is large enough, the rms quantities are almost equal to the $\sigma$ values.} 

\subsection{Bunched beam}

Ion bunches are formed along the accelerator circumference if the amplitude of the gap voltage is increased. Because the fundamental equation $f_{RF} = h\cdot f_R$, connecting the revolution frequency $f_R$ of the ions and the RF frequency $f_{RF}$, is valid, the harmonic number $h$ defines, how many bunches occur\footnote{Empty buckets will not be discussed in this paper.}. For small oscillation amplitudes with respect to the reference particle the ions follow elliptic trajectories in phase space. Hence the emittance of one single bunch, which is uniformly and continuously distributed and sufficiently short, can be approximated by
\begin{align}
A_{b,bunch}^{unif} &= \pi\cdot\Delta\hat{W}\cdot\Delta\hat{t}\,.\label{eq_A_b}
\end{align}

Again, $\Delta \hat{W}$ and $\Delta\hat{t}$ belong to the outermost particle in the bunch \marker{with respect to the energy deviation $\Delta W$ and the time deviation $\Delta t$} (cf.~\cite{hinterberger}).

In the case of a continuous Gaussian distribution, several emittance definitions exist. In analogy to transverse emittance definitions including a time-correlated energy spread, the emittance definition
\begin{align}
A_{b,bunch}^{Gauss} &\sim \sqrt{\sigma_{\Delta W}^2\cdot\sigma_{\Delta t}^2-\langle\Delta W\,\Delta t\rangle^2}\label{eq_A_b_sig3}
\end{align}
is commonly used (cf.~\cite{cornacchia,lee}). \marker{Here, $\sigma_{\Delta W}$ and $\sigma_{\Delta t}$ denote the $\sigma$ values of the energy deviation and the time deviation, respectively.} Neglecting the covariance term\footnote{The validity of this approximation is confirmed by our simulations.} $\langle\Delta W\,\Delta t\rangle$ directly leads to the next definition \eqref{eq_A_b_sig1}. Following equation \eqref{eq_A_b} the principal axes $\Delta\hat{W}$ and $\Delta\hat{t}$ are often replaced by the $\sigma_{\Delta W}$ and $\sigma_{\Delta t}$ values in phase space. Hence,
\begin{align}
A_{b,bunch}^{Gauss} &\sim \pi\cdot\sigma_{\Delta W}\cdot\sigma_{\Delta t}\,.\label{eq_A_b_sig1}
\end{align}
can be defined.

Of course, the area defined by the right hand side of equation \eqref{eq_A_b_sig1} does not contain 95\% of the particles if no constant factor is introduced. As stated in \cite{lee,syphers}, for a Gaussian distribution the area is given by
\begin{align}
A_{b,bunch}^{95\%}&= 6\cdot\pi\cdot\sigma_{\Delta W}\cdot\sigma_{\Delta t}\,,\label{eq_A_b_sig2}
\end{align}
if the 95\% definition is used.

As we already did for the coasting beam, we make use of the \marker{same} factor $k$ \marker{as in equation \eqref{eq_A_b,coast_sigma}}, since neither enclosing 95\% of the particles nor other definitions ensure emittance preservation. \marker{For each semi-axis of the phase space area with a 2D Gaussian distribution this factor $k$ is applied once.} Taking both phase space dimensions into account we obtain 
\begin{align}
A_{b,bunch}^{Gauss} &= k^2\cdot\pi\cdot\sigma_{\Delta W}\cdot\sigma_{\Delta t}\label{eq_A_b_sig4}
\end{align}
for a continuous distribution. 
Every $\sigma$ value, i.e.~every $\sigma_{\Delta W}$ and every $\sigma_{\Delta t}$, that occurs is multiplied with the factor $k$ once. This leads to a linear representation in equation (\ref{eq_A_b,coast_sigma}) and quadratic representation in equation (\ref{eq_A_b_sig4}).
\DB{The usage of the same factor in these equations results from the idea to find a definition for the rms emittance that ensures emittance preservation. Therefore, we introduce the same factor $k$ for every Gaussian distributed phase space dimension. It would also be possible to introduce three different factors ($k_1$, $k_2$, $k_3$) but that would neither lead to different results nor give additional information as we will show in section \ref{subsec_emit_pres}. }

Now we assume a discrete particle distribution which is approximately Gaussian.
Using rms quantities for the energy $\Delta W_{rms}$ and the time $\Delta t_{rms}$ according to
\begin{align}
\overline{\Delta t} &= \frac{1}{N}\sum\limits_{i=1}^N \Delta t_i\label{eq_t_mean}
\end{align}
and equation \eqref{eq_Wmean}, as well as
\begin{align}
\Delta t_{rms} &=\sqrt{ \frac{1}{N}\sum\limits_{i=1}^N \left(\Delta t_i-\overline{\Delta t}\right)^2}\label{eq_t_rms}
\end{align}
and equation \eqref{eq_Wrms} \marker{for an $N$-particle set with an energy deviation $\Delta W_i$ and a time deviation $\Delta t_i$ per particle $i$, we define the rms emittance of a bunch by}
\begin{align}
A_{b,bunch} &= k^2\cdot\pi\cdot\Delta W_{rms,bunch}\cdot\Delta t_{rms,bunch}\label{eq_A_b,bunch_rms}\,.
\end{align}

\marker{In the following sections the rms definitions \eqref{eq_A_b,coast_rms} and \eqref{eq_A_b,bunch_rms} are used for evaluating the phase space data sets of the simulations.
	
Because $k$ occurs quadratically in equation \eqref{eq_A_b,bunch_rms} but only linearly in equation \eqref{eq_A_b,coast_rms}, we are able to determine $k$ in a unique way as we will show.
	
\subsection{Emittance preservation}
\label{subsec_emit_pres}
As mentioned before emittance preservation is guaranteed by Liouville's theorem for continuous beams (cf.~\cite{lee,garoby}). Furthermore, the occupied phase space area remains constant if a coasting beam is bunched provided a sufficiently adiabatic capture process is performed. Under this condition, we now require preservation for our rms emittance definitions \eqref{eq_A_b,coast_rms} and \eqref{eq_A_b,bunch_rms}, which is possible because the value of $k$ is not yet fixed. As a result, 
\begin{align}
A_{b,coast} &\stackrel{!}{=}h\cdot A_{b,bunch}\label{eq_emitt}\\
&\Rightarrow\,2\cdot k\cdot\Delta W_{rms,coast}\cdot T_R\notag\\
&\stackrel{!}{=}h\cdot k^2\cdot\pi\cdot\Delta W_{rms,bunch}\cdot\Delta t_{rms,bunch}\label{eq_liou}
\end{align}  
is required. Thus \marker{our introduced} factor
\begin{align}
k&=\frac{2\cdot\Delta W_{rms,coast}\cdot T_R}{h\cdot\pi\cdot\Delta W_{rms,bunch}\cdot\Delta t_{rms,bunch}}\label{eq_k_rms}
\end{align} 
can be calculated by means of the rms quantities \marker{$\Delta W_{rms,coast}$, $\Delta W_{rms,bunch}$ and $\Delta t_{rms,bunch}$} if the initial and final states of the bunching process are known. \marker{Again, $T_R$ denotes the revolution time of the reference particle and $h$ the harmonic number.}

\DB{Defining three different parameters ($k_1$, $k_2$, $k_3$) leads to the same result. Equation \eqref{eq_liou} then becomes 
\begin{align}
&2\cdot k_1\cdot\Delta W_{rms,coast}\cdot T_R\notag\\
&\stackrel{!}{=}h\cdot k_2 \cdot k_3 \cdot\pi\cdot\Delta W_{rms,bunch}\cdot\Delta t_{rms,bunch}
\end{align}  
which leads to
\begin{align}
k_4\stackrel{!}{=} \frac{k_2 \cdot k_3}{k_1}&=\frac{2\cdot\Delta W_{rms,coast}\cdot T_R}{h\cdot\pi\cdot\Delta W_{rms,bunch}\cdot\Delta t_{rms,bunch}}.
\end{align} 
This factor $k_4$ has the same definition as $k$ and therefore contains the same information. Since there is no further reason to assume that the factors are different and all distributions are Gaussian (i.e. they are distributed in the same way), we will use the factor $k$ below as given in equation \eqref{eq_k_rms}. This way, after determination of $k$, equations \eqref{eq_A_b,coast_rms} and \eqref{eq_A_b,bunch_rms} give rms emittance definitions that ensure emittance preservation in our case. This would not be possible for the usage of differing factors since we then do not know how the value of $k_4$ distributes among $k_1$, $k_2$ and $k_3$.
}

\section{Simulation setup}

The bunching process in a heavy-ion synchrotron is simulated in order to confirm the theoretical emittance considerations of the previous sections. Thereby, a coasting beam is bunched by means of a linear increase of the gap voltage within a given ramp time. 

\marker{Due to time constraints iso-adiabatic ramps are used in practice, which are faster than linear ramps. But these iso-adiabatic ramps always lead to an emittance blow-up even if it is small, since they do not start at a zero gap voltage. For that very reason, linear ramps that start at zero and can therefore be expected to fulfill emittance preservation for an infinitely long time were investigated.} 
Simulations with longer and longer ramp time are carried out to guarantee that bunching is happening slowly enough to ensure adiabaticity. 

In the beginning, the Gaussian distribution of the coasting beam was verified by calculating the rms energy deviation $\Delta W_{rms,coast}^{sim}$ according to equations \eqref{eq_Wmean} and \eqref{eq_Wrms} of the start ensemble and $\Delta W_{\sigma,coast}^{sim}$ by a Gaussian fit of its projection onto the energy axis. The resulting momentum spreads show a negligible deviation from the simulation input parameter (error of $\frac{\Delta p_{rms,coast}^{sim}}{p_R}\big|_{\sigma}<\unit[0.3]{\%}$ and error of $\frac{\Delta p_{\sigma,coast}^{sim}}{p_R}\big|_{\sigma}<\unit[0.7]{\%}$). Therefore, it is confirmed that the pseudo random number generator in the simulation software does not lead to significant errors.

TABLE~\ref{simulationParameters} lists the simulation parameters based on \marker{those of the SIS18 accelerator at GSI} and the ion Rapid Cycling Medical Synchrotron (iRCMS) proposed in \cite{iRCMS}.

\begin{table}[htb]
	\caption{Simulation parameters \label{simulationParameters}}
	\begin{ruledtabular}
		\begin{tabular}{lllll}
			Circumference $C$ \DB{(m)} & $216.72$ and $60$ \\ [6pt]
			Transition gamma $\gamma_t$ & $5.45$ and $4.207$ \\ [6pt]
			Ion species & p, $^{238}\mathrm{U}^{73+}$ and $^{40}\mathrm{Ar}^{18+}$\\[6pt]
			Number of macro particles $N$ & $20 \DB{\times} 10^3$, $50 \DB{\times} 10^3$ and $180\DB{\times} 10^3$\\[6pt]
			Kinetic energy $W_{kin}$ \DB{(MeV/u)} & $11.21$ and $400$ \\[6pt]
			Momentum spread $\frac{\Delta p}{p}\big|_{\sigma}$ & $1 \DB{\times} 10^{-4}$ and $5 \DB{\times} 10^{-4}$ \\[6pt]
			Harmonic number $h$ & $1$ and $4$ \\[6pt]
			Ramp time $t_{ramp}$ \DB{(ms)} & $100-1400$ \\[6pt]
			Final gap voltage $\hat V_{max}$ \DB{(kV)} & $10$ and $32$ \\
		\end{tabular}
	\end{ruledtabular}
\end{table}

Several simulation scenarios with varying ion species and different values for the kinetic energy, the momentum spread, target gap voltage and ramp time were performed. In addition, different accelerator layouts are taken into account by varying the synchrotron circumference and transition gamma.

The synchrotron RF simulator $SyncRF$ \marker{\cite{syncrf}} and the particle tracking code BlonD \cite{BlonD} are used to analyze the longitudinal beam dynamics taking no higher-order effects like beam loading, space charge or impedances into account. Large amounts of data were produced by simulation durations in the range of hours. 

\section{Simulation results}
The results presented in the following are provided by both tracking codes. This serves as additional cross-check. 

\subsection{Gaussian shape of the bunched beam current}
\begin{figure}[htb]
	\centering
	\includegraphics[width=0.5\textwidth]{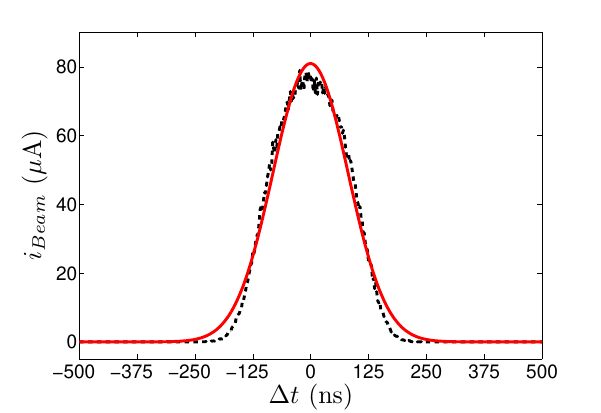}
	\caption{Beam current of the bunched proton beam (p at a kinetic energy $W_{kin}=\unit[400]{MeV}$, $h=1$, $\hat{V}_{max}=\unit[32]{kV}$, $\frac{\Delta p}{p_R}\big|_{\sigma}=5 \DB{\times} 10^{-4}$): Simulation (dashed black) and Gaussian fit (solid red).}
	\label{fig_prj_fit}
\end{figure}
The simulations confirmed that a Gaussian distribution of the coasting beam momentum spread leads to a beam current shape of the bunched beam that can satisfactorily be described by a Gaussian distribution. FIG.~\ref{fig_prj_fit} shows an example for the beam current (dashed black) obtained by a projection of a tracking simulation ensemble (with 180000 macro particles and 450 projection bins at the harmonic number $h$=1) if the coasting beam is bunched in \unit[800]{ms} and the corresponding Gaussian fit (solid red). \DB{The deviation between the Gaussian fit and the obtained beam profile is smaller than the accuracy at which we are aiming in this paper. Therefore, different distributions are not discussed here.}

\subsection{Emittance preservation factor $k$}
Now, the unknown parameter $k$ shall be calculated according to equation (\ref{eq_k_rms}). Since the finite ramp
time $t_{ramp}$ in the simulation will always lead to a non-ideal situation, only a factor
\begin{equation}
\label{factorKtilde}
\tilde k = \frac{2 \cdot \Delta W_{rms, coast}^{sim} \cdot T_R}{h \cdot \pi \Delta W_{rms, bunch}^{sim} \cdot \Delta t_{rms, bunch}^{sim}}
\end{equation}\\
with $\Delta W_{rms, coast}^{sim}$, $\Delta W_{rms, bunch}^{sim}$ and $\Delta t_{rms, bunch}^{sim}$ as the rms values of the ensemble data can be calculated for each specific simulation. For the ideal case $t_{ramp} \rightarrow \infty$, one can expect $\tilde k \rightarrow k$.

\begin{figure}[htb]
	\includegraphics{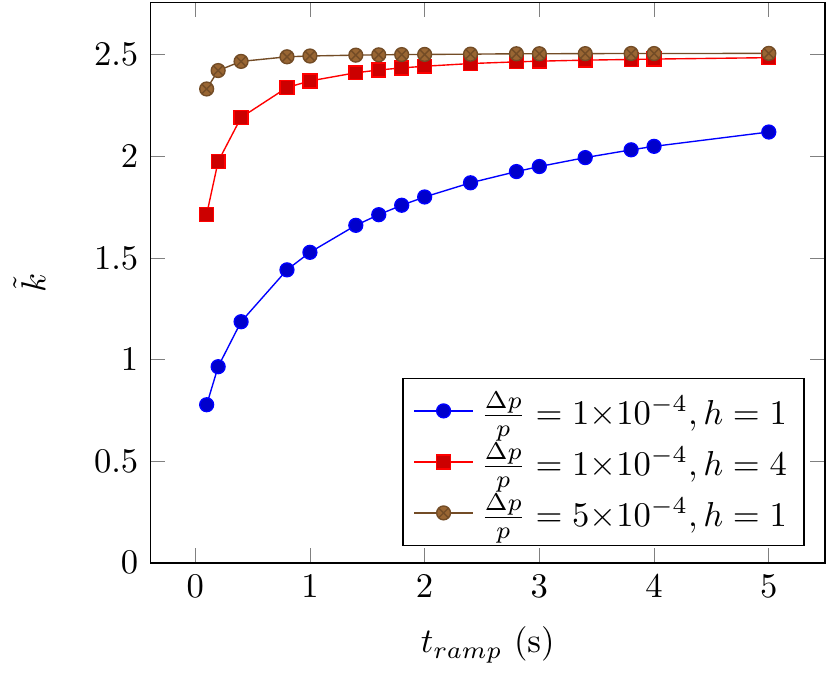}
	\caption{Emittance preservation factor $\tilde k$ of a bunched proton beam (p at a kinetic energy $W_{kin}=\unit[400]{MeV}$, $C=\unit[216.72]{m}$, $\hat{V}_{max}=\unit[32]{kV}$, $\gamma_t = 5.45$) for different harmonic numbers $h$ and momentum spreads $\frac{\Delta p}{p}\big|_{\sigma}$.}
	\label{k_over_time}
\end{figure}

Various combinations of parameters lead to significantly different bunching times to approach the limit value of $k$. The factors with the greatest influence are the harmonic number $h$ and the momentum spread $\frac{\Delta p}{p}\big|_{\sigma}$. If increased, both of them lead to the limit value being reached more quickly. FIG.~\ref{k_over_time} shows the emittance preservation factor $\tilde k$ plotted over the ramp time for different parameter sets to illustrate the different bunching times but all curves are heading to $\tilde k \approx 2.5$. 
Equivalent results are produced for all other parameter combinations mentioned in TABLE \ref{simulationParameters} except for those discussed in section \ref{subsec_limits}.

The simulation results suggest that $\tilde k \rightarrow k = 2.5$ is valid for all scenarios if the ramp time is increased. This implies that there is a fixed ratio between the rms emittance before and after the adiabatic process. Assuming the general validity of $k=2.5$, the rms emittance after the bunching process can be calculated if the rms emittance of the coasting beam is known. One can even determine $\Delta W_{rms, bunch}$ and $\Delta t_{rms, bunch}$ individually. For this purpose one can make use of the axis ratio 
\begin{equation}
\frac{\Delta W_{rms, bunch}}{\Delta t_{rms, bunch}}=f_s \frac{2\pi W_R \beta_R^2}{|\eta_R|}
\end{equation}
from \cite{Klingbeil2015} with $f_s$ as the synchrotron frequency, $\eta_R$ the phase slip factor and the relativistic factor $\beta_R$.

In order to verify \DB{the} convergence of $\tilde k$ to this value of $k = 2.5$, an error $\theta$ and a normalization of the ramp time is introduced in the next section.

\subsection{Convergence analysis}
The relative error between simulation and ideal value is given by
\begin{equation}
\theta = \frac{k-\tilde k}{k}
\end{equation}
and can be analyzed with respect to the time of the bunching process to guarantee emittance preservation. 

The normalized ramp time $\tilde t_{ramp}$ can be written as 
\begin{equation}
\tilde t_{ramp} = \frac{t_{ramp}}{\tilde T}
\end{equation}
with 
\begin{equation}
\tilde T = \frac{\hat V_{max} \cdot z \cdot C}{h^2 \cdot \Bigl(\frac{\Delta p}{p}\big|_{\sigma} \Bigr)^3 \cdot A_i} \cdot \DB{1\mathrm{s}\cdot 1\mathrm{V}^{-1}\cdot 1\mathrm{m}^{-1}}.
\end{equation}
Here $z$ denotes the charge number and $A_i$ the mass number. The normalization factor $\tilde T$ was obtained by considerations concerning fundamental properties of the tracking equations and empirical convergence analyses by means of the variety of simulations. This causes all simulation results with the same transition gamma and the same kinetic energy to lie on the same convergence curve.

FIG.~\ref{convergence_all} shows the error $\theta$ as a function of the normalized bunching time $\tilde t_{ramp}$ for $\gamma_t = 5.45$. Since $W_{kin}$ does not appear in $\tilde T$, two curves are shown. Both curves consist of several curves with different parameter combinations from TABLE \ref{simulationParameters} plotted with the same color to better illustrate the convergence behavior. Similar visualizations have been created for $\gamma_t = 4.207$.

\begin{figure}[htb]
	\includegraphics{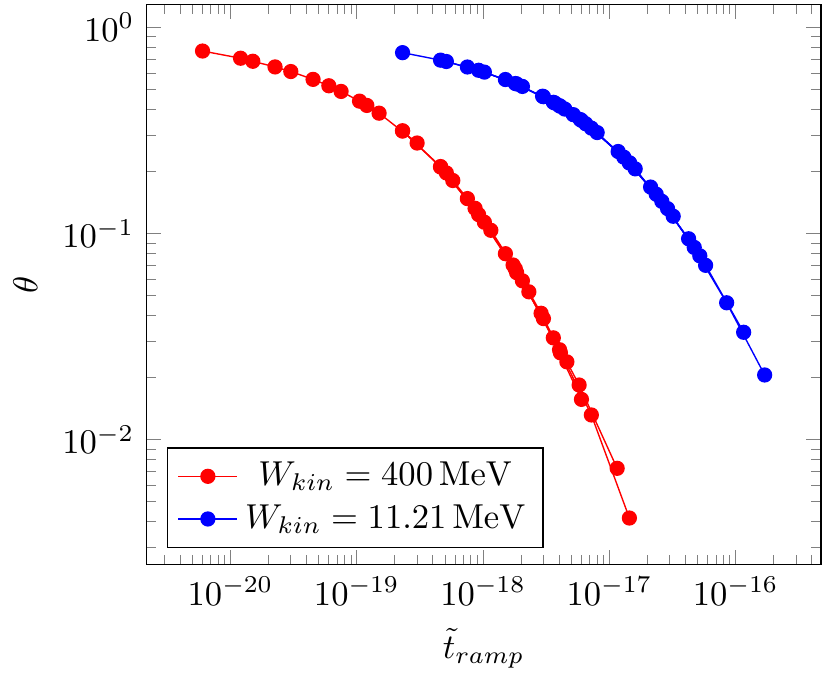}
	\caption{Error $\theta$ as a function of the normalized bunching time $\tilde t_{ramp}$ for $\gamma_t = 5.45$ and different energies. Both plots consist of several curves with different parameter combinations from TABLE~\ref{simulationParameters} plotted with the same color.}
	\label{convergence_all}
\end{figure}

As shown in FIG.~\ref{convergence_all}, the normalization leads to a convergence behavior that is much more transparent. Depending on the bunching process parameters, emittance preservation is only guaranteed for a sufficiently long ramp time. Nevertheless, FIG.~\ref{convergence_all} clearly shows that the factor $\tilde k$ converges to
\begin{equation}
k = 2.5.
\end{equation}
This holds for all bunching scenarios except for those in section \ref{subsec_limits}.

The longer the bunches, the faster the emittance preservation is ensured, i.e.~a low target gap voltage, a large harmonic number, a higher energy, a higher momentum spread of the coasting beam or a higher mass-to-charge ratio lead to a smaller error. The results confirm that the definition of a common factor $k$ is valid. Please note that according to equations \eqref{eq_A_b,coast_rms} and \eqref{eq_A_b,bunch_rms}, this definition leads to different percentages of enclosed particles for the coasting beam ($98.76  \%$) and for the bunched beam ($95.61 \,\%$). Thus, a unified emittance definition for both, a coasting beam and a bunched beam connected by the factor $k$, based on Liouville's theorem is proposed here.

\subsection{Limitations of the emittance preservation factor}
\label{subsec_limits}
A convergence behaviour as in FIG.~\ref{convergence_all} can be observed up to an error of $0.3 \,\%$. A further increase of the bunching time does not lead to smaller errors but to numerical fluctuations. Nevertheless, this error is small enough to justify the introduced factor convincingly.

The presented factor $k$ can no longer be reliably applied if the final gap voltage is too low to capture the entire bunch. 
This result is not surprising, since the approximation of an elliptical phase space area is no longer permissible in this case and thus the computation of the emittance as done in this paper is no longer valid. Furthermore, this parameter combination is not a constellation that would be used in a real accelerator. This happens for parameter combinations with $\hat V_{max}=\unit[10]{kV}$ and $\frac{\Delta p}{p}\big|_{\sigma}=5 \DB{\times} 10^{-4}$. A gap voltage of $\unit[10]{kV}$ is not able to capture the whole bunch in that case as the tracking simulations show.
Furthermore, the derived factor $k$ is not able to make a statement about how long the adiabatic process lasts. To do this, the kinetic energy and the transition gamma would have to be introduced into the normalization factor $\tilde T$. 

\DB{Besides, a theoretical demonstration of $k = 2.5$ is lacking. This should be the focus of future work. Since we expect a high mathematical effort, this is not relevant for this paper. Furthermore, in this publication we do not state any mathematical strictness of the found numerical value but point to a connection we have found which can be confirmed by simulations up to the accuracy computed by us.}

\section{Summary}
This contribution presents a specific definition of the longitudinal rms emittance connecting the rms emittances of a coasting and a bunched ion beam by the factor $k$ (cf.~eqn.~(\ref{eq_k_rms})). This definition is motivated by Liouville's theorem and adiabaticity considerations. By means of the analysis of various simulation scenarios of an ion beam bunching processes, the factor was determined to be
$k = 2.5$. The usefulness of this factor results from the fact that it can replace a complex simulation. It provides the possibility to specify the resulting bunch for a given coasting beam without performing a time-consuming simulation as long as the bunching process is sufficiently adiabatic. Since the emittance of the bunched beam is known through the factor $k$, it is also possible to determine the required bucket area and thus the final gap voltage needed to capture the bunch for a specific filling ratio of the bucket.

\bibliography{heavy_ion_bunching}
\end{document}